\documentclass[pdflatex,sn-mathphys-num]{sn-jnl}
\DeclareUnicodeCharacter{2009}{\,}
\usepackage{graphicx}%
\usepackage{multirow}%
\usepackage{amsmath,amssymb,amsfonts}%
\usepackage{amsthm}%
\usepackage{mathrsfs}%
\usepackage[title]{appendix}%
\usepackage{xcolor}%
\usepackage{textcomp}%
\usepackage{manyfoot}%
\usepackage{booktabs}%
\usepackage{subcaption}%
\usepackage{algorithm}%
\usepackage{algorithmicx}%
\usepackage{algpseudocode}%
\usepackage{listings}%
\usepackage{gensymb}
\newcommand\arcmin{\mbox{$^\prime$}}%
\newcommand\arcsec{\mbox{$^{\prime\prime}$}}%

\theoremstyle{thmstyleone}%
\theoremstyle{thmstyletwo}%

\theoremstyle{thmstylethree}%

\begin{document}

\title[Article Title]{Atmospheric Dispersion Measurement at Hanle Site}


\author*[1,2]{\fnm{Manjunath} \sur{Bestha}}\email{bestha95@gmail.com}

\author[1]{\fnm{Sivarani} \sur{Thirupathi}}\email{sivarani@gmail.com}
\equalcont{These authors contributed equally to this work.}

\author[3]{\fnm{Athira} \sur{Unni}}\email{athira.exo@gmail.com}
\equalcont{These authors contributed equally to this work.}

\author[1,2]{\fnm{Parvathy} \sur{M}}\email{parvathy.m@iiap.res.in}
\equalcont{These authors contributed equally to this work.}

\author[4]{\fnm{Devika} \sur{Divakar}}\email{devika.divakar@austin.utexas.edu}
\equalcont{These authors contributed equally to this work.}


\affil*[1]{\orgname{Indian Institute of Astrophysics}, \orgaddress{\city{Bengaluru}, \country{India}}}

\affil[2]{\orgname{Calcutta University}, \orgaddress{\city{Kolkata}, \country{India}}}

\affil[3]{\orgname{University of California}, \orgaddress{\city{Santa Cruz}, \state{California}, \country{USA}}}

\affil[4]{\orgname{University of Texas}, \orgaddress{\city{Austin}, \state{Texas}, \country{USA}}}


\abstract{
Atmospheric dispersion introduces wavelength-dependent effects that significantly impact ground-based observations, particularly in slit- and fibre-fed spectroscopic studies. These effects reduce the signal entering the spectrograph and introduce systematic errors in radial velocity measurements. To address this challenge, atmospheric dispersion correctors are utilised. However, many existing designs of these correctors, which are based on theoretical models, often lack practical validation and consistency.
The forthcoming National Large Optical Telescope (NLOT) will be installed at Hanle, a site known for its favourable astronomical sky conditions. Thus, the design of an effective dispersion corrector for the instruments on the NLOT, specifically one that compensates for the measured dispersion, is crucial. For the first time, we have directly measured atmospheric dispersion at the Hanle site using the Himalayan Faint Object Spectrograph mounted on the Himalayan Chandra Telescope. In this study, we present our methodology, the dispersion measurements obtained within the 400 to 700 nm wavelength range, and a comparison with modelled dispersion values.

}


\maketitle

\section{Introduction}\label{sec1}

The Indian Astronomical Observatory (IAO) is situated in Hanle, Ladakh, at a latitude of 32.7797\degree,  and longitude of 78.9639\degree, at an altitude of approximately 4500 meters, and benefits from low atmospheric pressure, minimal humidity, and reduced light pollution. These factors create an optimal environment for precision astronomical spectroscopy and photometric observations.


Recognising these favourable characteristics, India has proposed the installation of the National Large Optical Telescope (NLOT), a 10-meter-class telescope, at Hanle \cite{Anupama2022}. The facility is expected to incorporate advanced instruments covering the wavelength range from 300nm in the blue to mid-IR wavelengths. Fiber-fed seeing-limited and adaptive optics-assisted are proposed, including high-stability instruments for detecting exoplanets through the radial velocity (RV) method and characterizing planetary atmospheres using high-dispersion spectrographs, which have been demonstrated at Hanle using a high and low resolution spectrograph \cite{Manju_HESP,Unni2024}. However, achieving high precision in RV measurements is contingent on several factors, including atmospheric dispersion.

To mitigate atmospheric dispersion effects, correctors have been designed \cite {Manju_Dispersion}. However, existing models used in their design tend to underestimate dispersion, particularly in the blue spectral region \cite{ZEMAX_MODEL_COR}, leading to less precise corrections. Consequently, onsite dispersion measurements are essential for refining corrector designs and improving observational accuracy. Future instrumentation advancements may focus on incorporating closed-loop correction systems to dynamically address dispersion-related challenges and enhance the overall precision of astronomical studies conducted at Hanle.

To address this issue, we have conducted on-site dispersion measurements using the Himalayan Faint Object Spectroscopic Camera (HFOSC), which is mounted on the Cassegrain unit of the 2-meter Himalayan Chandra Telescope (HCT) at the Indian Astronomical Observatory (IAO). The observational strategy, measurement methodology, and results obtained from this study are presented in this paper.

\section{Observations}

\begin{figure}[htbp]
    \centering
    \begin{subfigure}{0.49\textwidth}
        \centering
        \includegraphics[width=\textwidth]{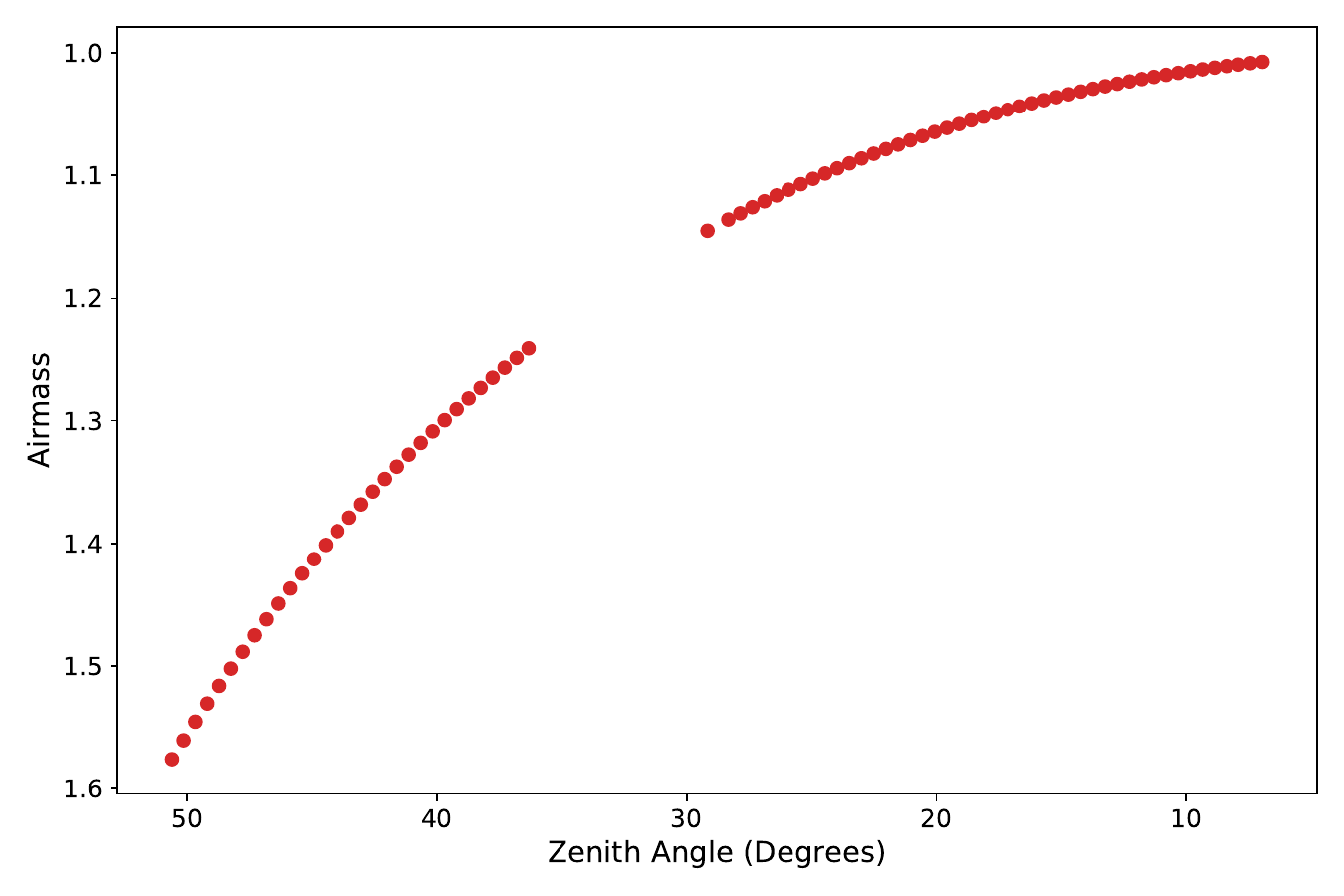}
        \caption{Airmass as a function of zenith angle for all frames of the observation.}
        \label{airmasss}
    \end{subfigure}
    \hfill
    \begin{subfigure}{0.49\textwidth}
        \centering
        \includegraphics[width=\textwidth]{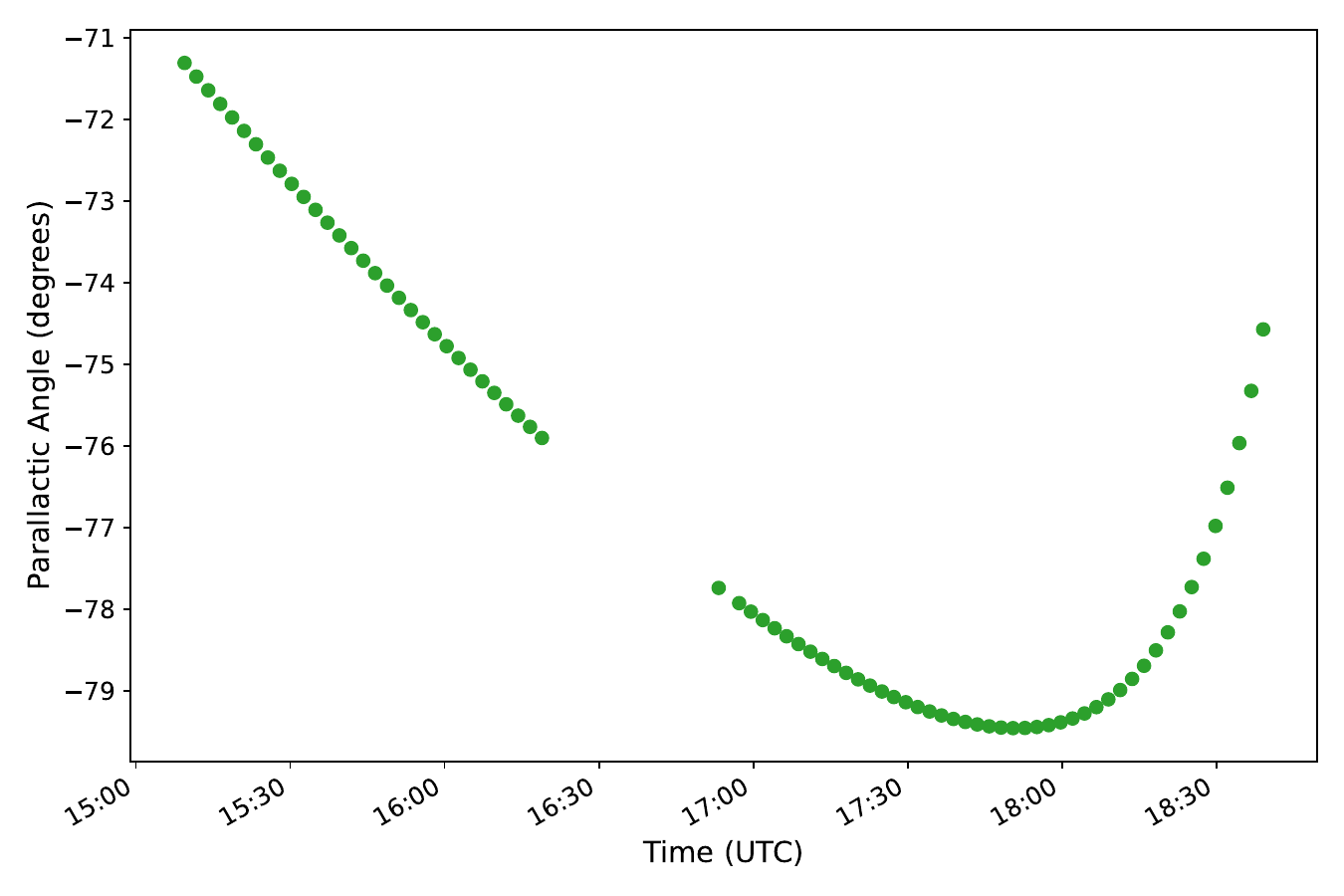}
        \caption{Parallactic angle as a function of elevation for all frames of the observation.}
        \label{paralactic_angle}
    \end{subfigure}
    \caption{Airmass and parallactic angle trends during the observation.}
    \label{dispersion_plots}
\end{figure}

On 9 July 2024, nearly 76 low-resolution optical spectra of KELT-20, a  V=8 magnitude star (RA: 294.66142\degree, Dec: 31.21920\degree), were obtained using HFOSC on the HCT. Each exposure lasted for 30 seconds, and the observations spanned a range of zenith angles from approximately 50\degree to 6\degree. Since the telescope has an alt-azimuth mount, observations could not be made at the zenith (0°) due to the extremely high apparent sky rotation at that position. To measure the atmospheric dispersion, the slit was aligned along the parallactic angle (Atmospheric dispersion direction), which has a rate of change that is equal to the rate of change in the field rotation during the observations. A wide slit 1340 µm (width: 15.41\arcsec, length: 11\arcmin) was used with Grism 7, which provides spectral coverage from 0.35 µm to 0.8 µm. Due to a target-of-opportunity observations program, there is a gap of approximately one hour in the data. The airmass and parallactic angle of each frame as a function of Zenith is shown in the Figure \ref{airmasss} \& \ref{paralactic_angle}.

\section{Methodology}
\begin{figure}[htbp]
    \centering
    \begin{subfigure}{0.49\textwidth}
        \centering
        \includegraphics[width=\textwidth]{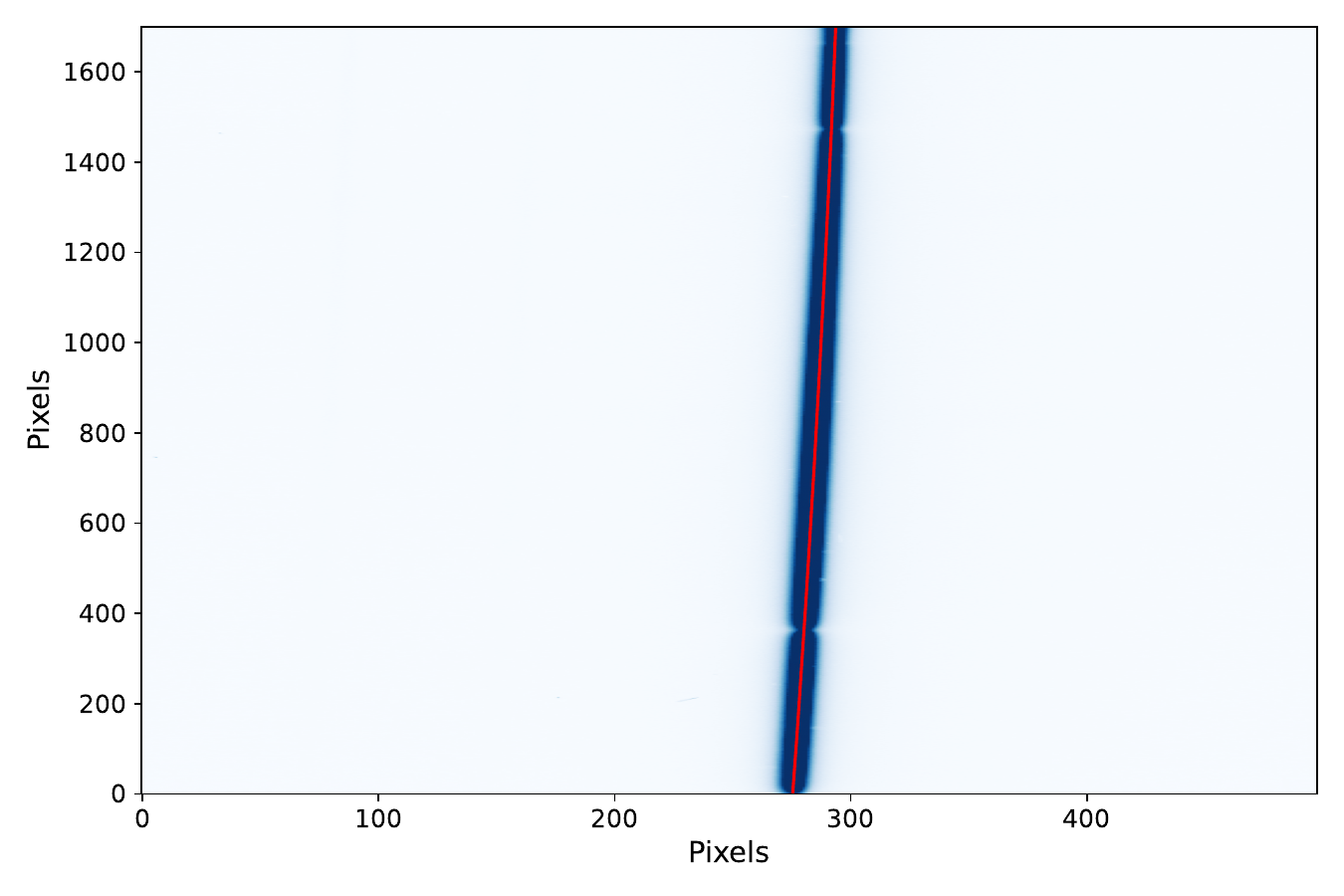}
        \caption{2D spectrum on the HFOSC detector, with the red line indicating the trace.}
        \label{2d_Image}
    \end{subfigure}
    \hfill
    \begin{subfigure}{0.49\textwidth}
        \centering
        \includegraphics[width=\textwidth]{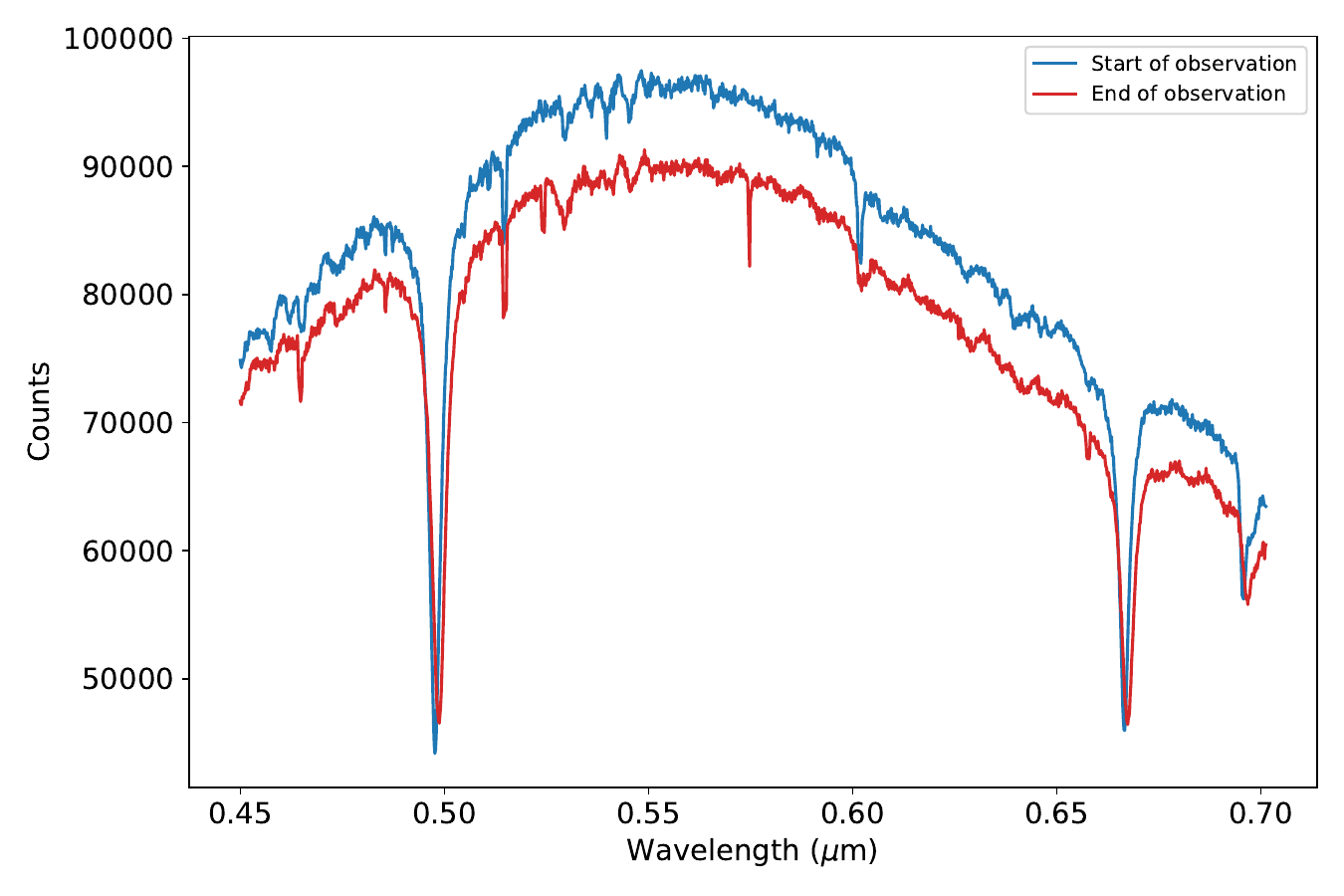}
        \caption{Extracted spectra of the target at the beginning and end of the observation.}
        \label{spectrum}
    \end{subfigure}
    \caption{Typical 2D image on the HFOSC detector and the extracted spectrum of the target.}
    \label{detector_image_and_spectrum}
\end{figure}

The methodology used to measure the dispersion is similar to that presented in Skemer et al.\citep{Skemer_2009}, except for the algorithm employed to determine the spectral trace, which is described below. 
Data reduction was performed using a custom Python-based pipeline built using the \texttt{PyKosmos} \cite{PyKosmos} and \texttt{CCDProc} modules. Standard reduction steps included bias subtraction, cosmic ray removal, and spectral trace identification, which was carried out by dividing the image into 10 bins, as larger bins tend to be influenced by strong stellar lines. Spectral extraction was performed by selecting 8 pixels on either side of the trace, corresponding to a total width of 16 pixels, based on the FWHM measured along the spatial direction at the center of the first image. (see Figure \ref{2d_Image} \& \ref{spectrum}). For wavelength calibration, we used \texttt{IRAF}, employing arc lamp spectra obtained with the FeAr (Iron-Argon) source for Grism 7.

For dispersion measurement, the trace plays a crucial role. In this study, we specifically focused on the trace shift in the spatial direction as a function of wavelength. This wavelength-dependent trace shift is primarily caused by a combination of atmospheric differential refraction and instrumental effects, such as field curvature, optical distortion, and mechanical flexure within the instrument. These instrumental artifacts become particularly significant near the edges of the detector.

\begin{figure*} 
    \centering
    \includegraphics[width=\textwidth]{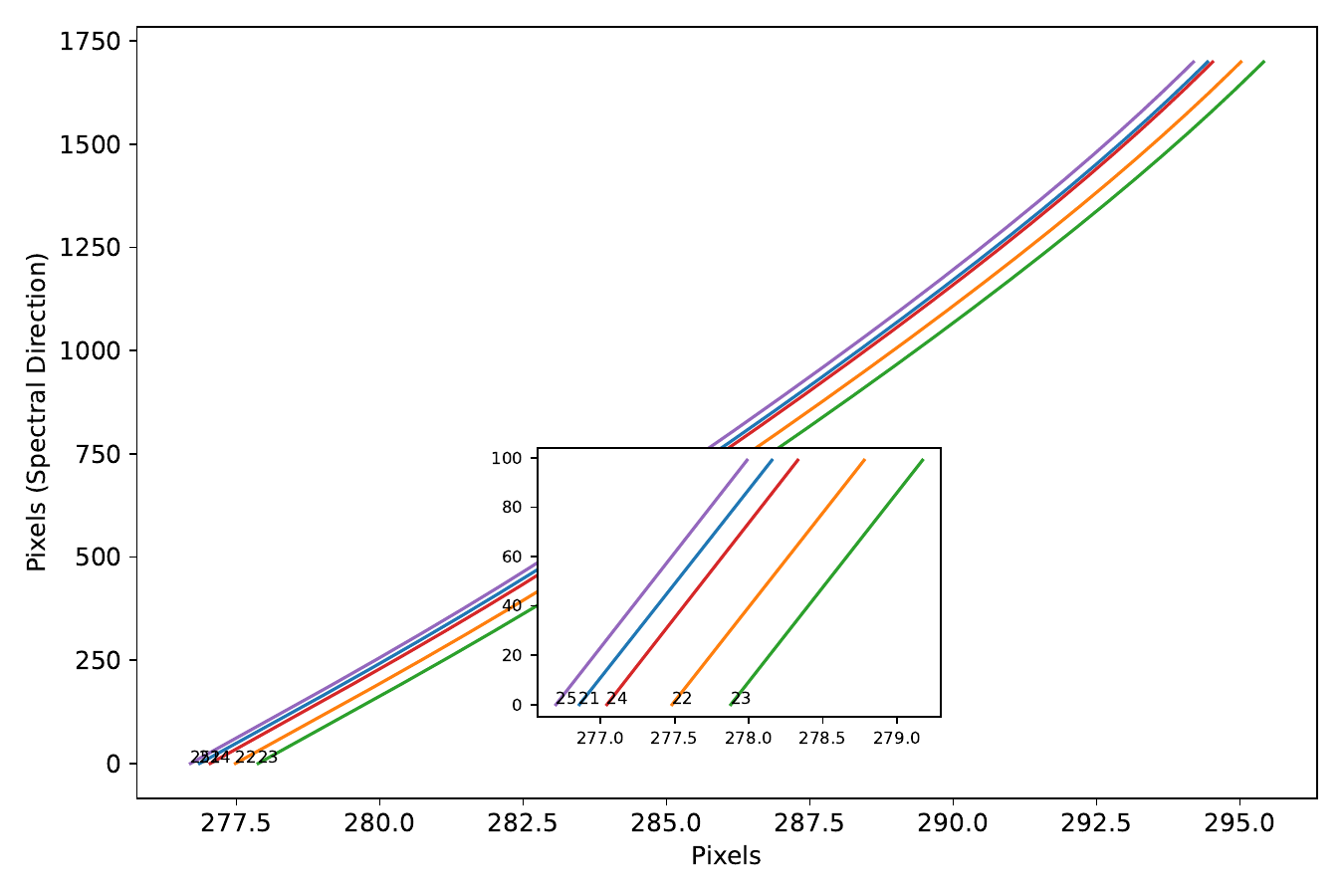}
    \caption{Example traces demonstrating random motion on the HFOSC detector during the observation.}
    \label{random_traces}
\end{figure*}

To mitigate these effects, we focused our analysis on the central region of the trace on the detector, specifically within the wavelength range of 0.4–0.7$\mu$m. We observed that the traces exhibited random shifts across the detector throughout the observation sequence. A few of these traces are shown in Figure \ref{random_traces}. The cause of the random shifts observed in the trace remains unclear. Additional observations and flat-field exposures will be used to understand this behaviour and the associated variations in the instrumental profile.

\begin{figure}[htbp]
    \centering
    \begin{subfigure}{0.49\textwidth}
        \centering
        \includegraphics[width=\textwidth]{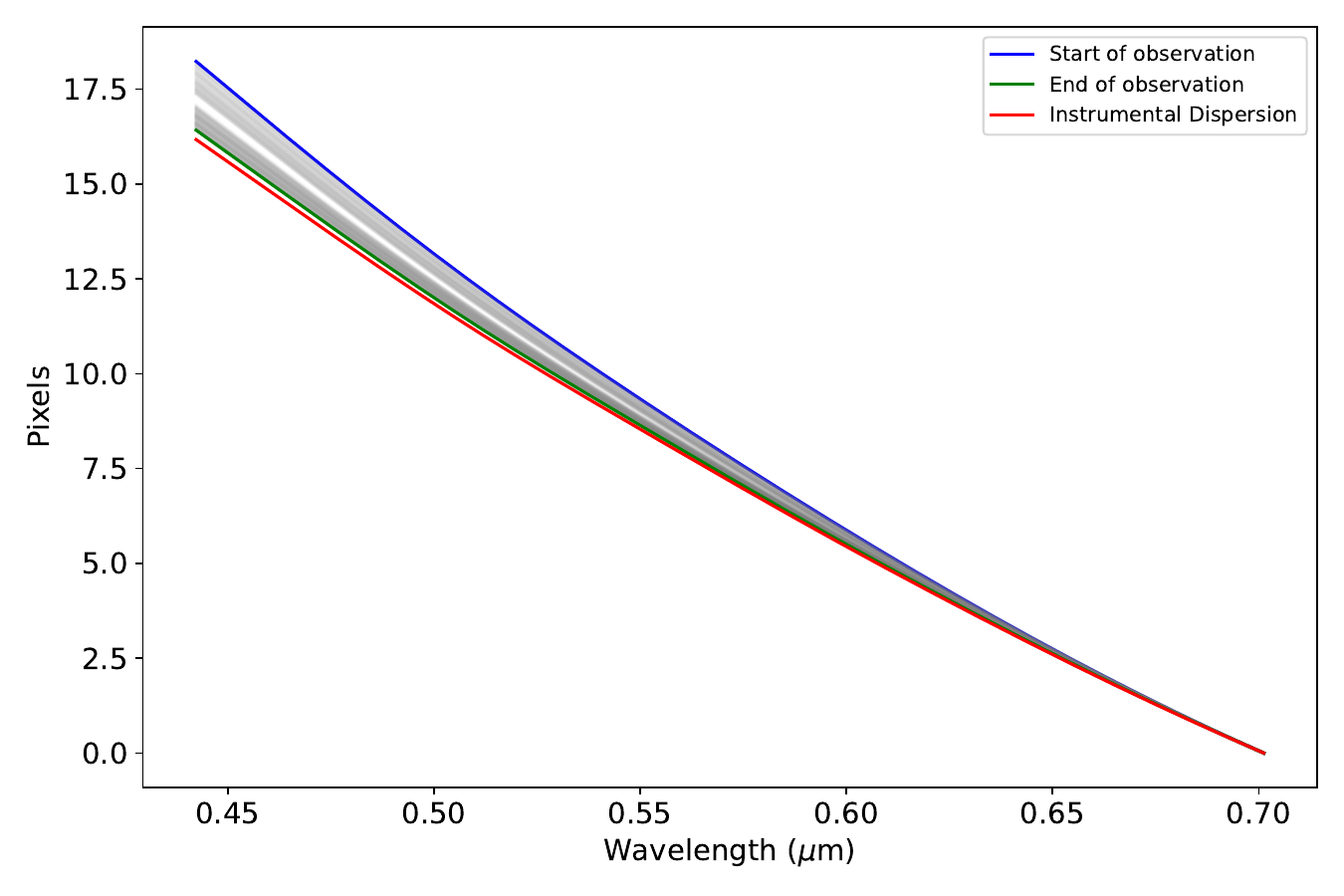}
    \caption{Traces aligned in the common frame (gray) and instrumental dispersion profile.}
        \label{traces}
    \end{subfigure}
    \hfill
    \begin{subfigure}{0.49\textwidth}
        \centering
        \includegraphics[width=\textwidth]{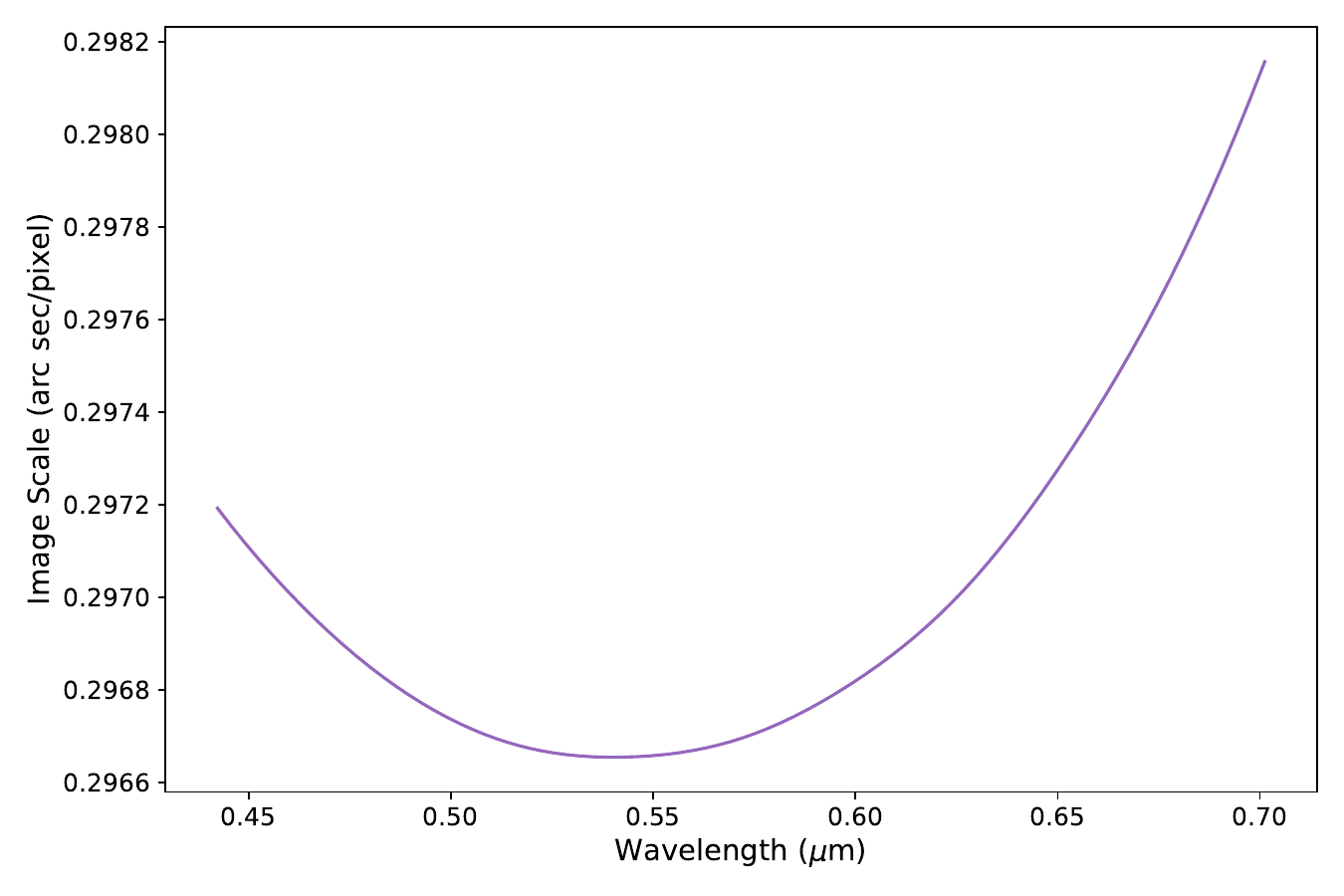}
        \caption{Pixel scale along the dispersion direction.}
        \label{pixel_scale}
    \end{subfigure}
    \caption{Traces in the common frame and pixel scale along the dispersion direction.}
    \label{traces_and_pixel_scale}
\end{figure}

For the current goal of estimating atmospheric dispersion,  we implemented a correction procedure by aligning all the traces to a common reference longer wavelength (0.7$\mu$m), where the dispersion is expected to be minimal. 
This normalization procedure was applied consistently to all traces, apparently bringing them to a common frame and allowing for coherent analysis across the dataset. Figure \ref{traces} shows the traces aligned in the common frame.  

In the Figure \ref{traces}, the distance between traces generally follows the expected trend with varying zenith angle along the wavelength axis, while at longer wavelengths, they do not show a clear trend with zenith angle. This could be due to a simple shift applied to correct for instrument drift. It is likely that the shape of the spectral trace might have changed over time. 

Afterwards, we determined the instrumental dispersion profile, which ideally should be measured at the zenith (0°), where atmospheric dispersion is negligible. However, since the Himalayan Chandra Telescope (HCT) is an alt-azimuth-mounted telescope, direct observations at the zenith were not feasible. As a result, we conducted observations at an altitude of 83°, corresponding to an airmass of approximately 1.007.

From the trace of this near-zenith observation, we subtracted the modelled atmospheric dispersion to isolate the instrumental component. The atmospheric dispersion was estimated using the Cassini atmospheric refraction model from the \texttt{AstroAtmosphere} \cite{AstroAtmosphere} module in Python, with input parameters including a median temperature of 285~K, relative humidity of 30\%, and atmospheric pressure of 590~mbar. This procedure allowed us to derive the instrumental dispersion profile (see Figure \ref{traces}).

Once this instrumental profile was determined, it was subtracted from the traces of all other science frames to isolate the atmospheric dispersion component.

\section{Results}
The measured atmospheric dispersion was initially in pixel units. To convert it into arc seconds, an accurate image scale is required. For this purpose, we used observational data from the same instrument configuration, where two targets (KELT-7 \& HD 241492) were aligned along the slit. This technique is commonly employed in ground-based, low-resolution transmission spectroscopy \cite{Unni2024}, where the angular separation between the two targets is known.

In our case, the actual separation between the two targets in the sky was 4.716 arc minutes. We measured the corresponding separation in pixels using the traces of both targets on the detector. By dividing the known angular separation (in arc seconds) by the measured pixel separation, we obtained the image scale in arc seconds per pixel along the spectral direction \cite{bachar} (see Figure \ref{pixel_scale}).

Finally, we converted the atmospheric dispersion from pixels to arc seconds by multiplying the pixel-based dispersion values by the calculated image scale.

\begin{figure}[htbp]
    \centering
    \begin{subfigure}{0.49\textwidth}
        \centering
        \includegraphics[width=\textwidth]{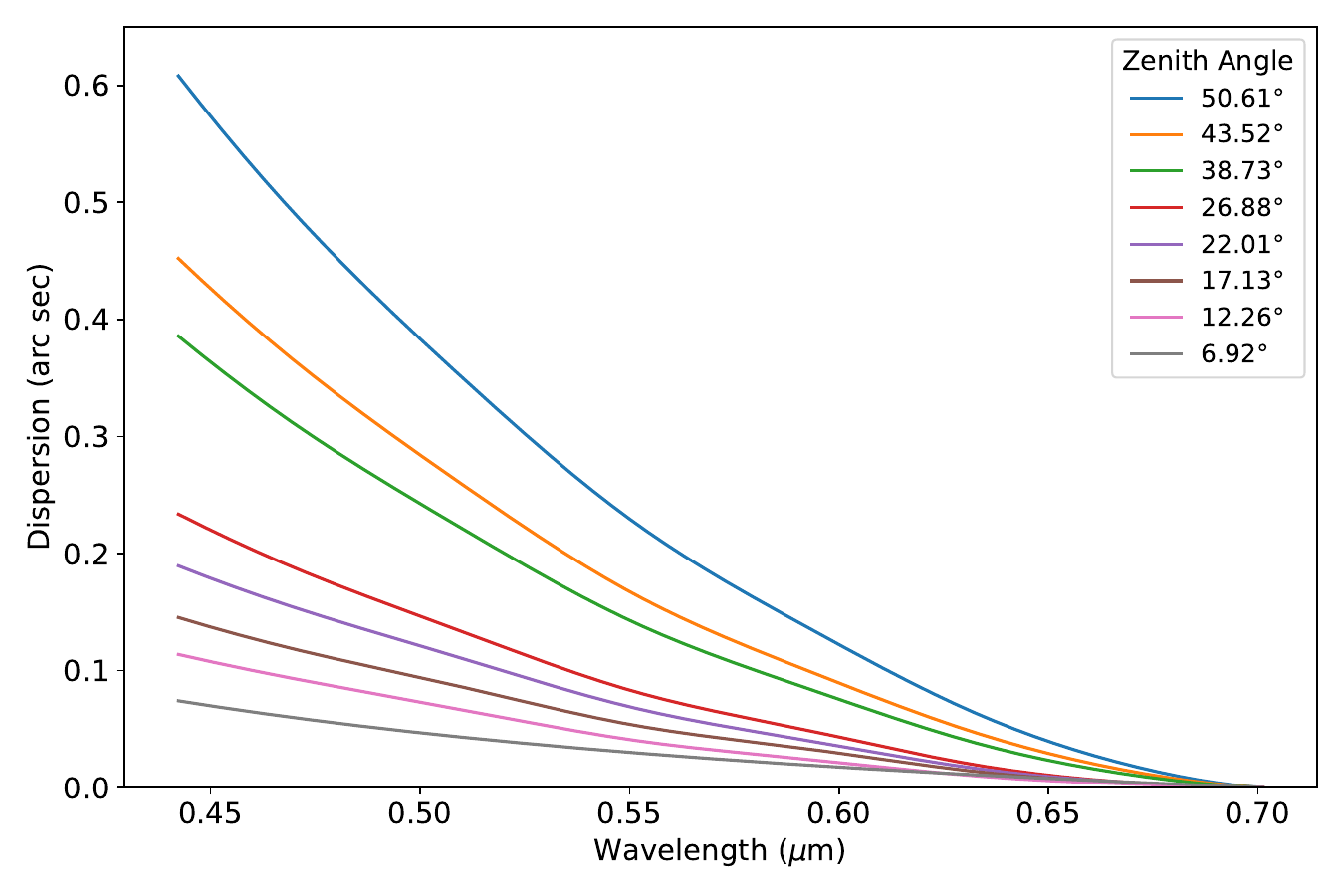}
        \caption{Measured atmospheric dispersion at the Hanle site.}
        \label{observed_dispersion}
    \end{subfigure}
    \hfill
    \begin{subfigure}{0.49\textwidth}
        \centering
        \includegraphics[width=\textwidth]{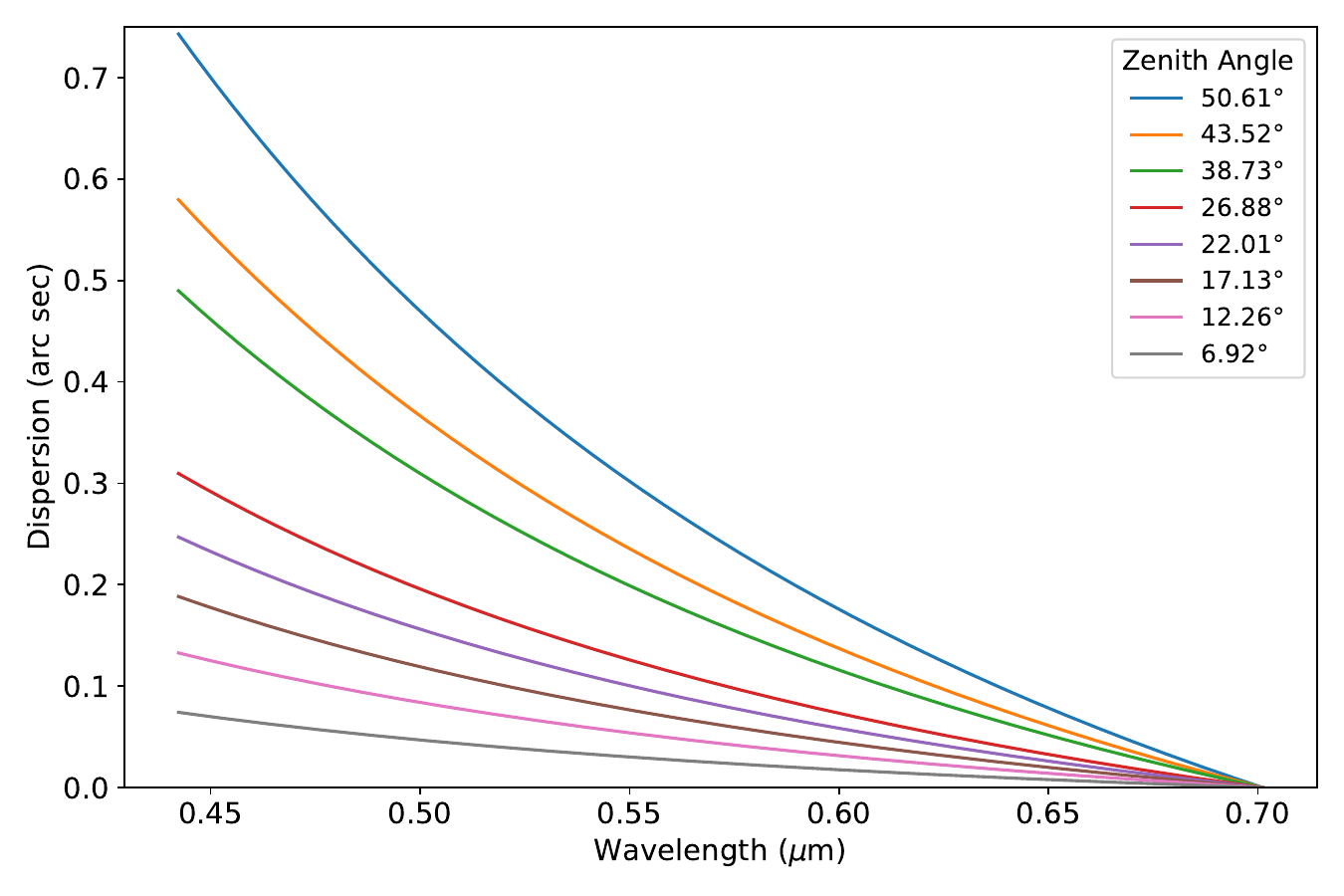}
        \caption{Modelled atmospheric dispersion using the Cassini model.}
        \label{modeled_dispersion}
    \end{subfigure}
    \caption{Comparison of measured and modelled atmospheric dispersion at different zenith angles.}
    \label{final_dispersion_plots}
\end{figure}

The atmospheric dispersion at various zenith angles is displayed in Figure \ref{observed_dispersion}. At a zenith angle of 50.61° at the Hanle site, the measured dispersion is 0.6082\arcsec using traces obtained by dividing the image into 10 bins. This value changes with the number of bins used, whereas the modelled dispersion is 0.7429\arcsec, as shown in Figure \ref{modeled_dispersion}. This results in a discrepancy of approximately 134.6 milliarcseconds. This difference may be attributed to the assumed median atmospheric parameters; even slight variations in temperature, pressure, or humidity can lead to deviations. Additionally, the discrepancy could be due to a slight underestimation of the instrumental dispersion profile, as we were unable to obtain observations at the zenith (0°), where atmospheric effects are minimal.

\section{Conclusion and Future Work}

In this work, we have measured atmospheric dispersion at the Hanle site with the aim of contributing to the development of dispersion correctors for high-stability instruments on the upcoming NLOT. The measured dispersion is 0.6082\arcsec at a zenith angle of 50.61°, which shows less than half a pixel deviation from the modelled dispersion. To verify and improve the accuracy of our results, we plan to incorporate the temperature, pressure, and humidity values recorded during each exposure of the observation. Additionally, we will include uncertainty estimates in the measurements. As a further validation step, we intend to cross-check our methodology using data obtained from telescopes with equatorial mounts.

\bmhead{Acknowledgments}

We thank the staff of the Indian Astronomical Observatory (IAO), Hanle, and the Center for Research \& Education in Science \& Technology (CREST), Hosakote, for their support in making these observations possible. The facilities at IAO and CREST are operated by the Indian Institute of Astrophysics (IIA), Bangalore. We also thank the HCT Time Allocation Committee for granting observing time for this project.

\bibliography{sn-bibliography}

\end{document}